\documentclass[draft]{elsart}
\usepackage{amssymb,amsmath}
\def\openone{\leavevmode\hbox{\small1\kern-3.8pt\normalsize1}}%
\def\ket#1{\vert {#1} \rangle}%
\def\kket#1{\vert {#1} \rangle\!\rangle}%
\def\bra#1{\langle {#1} \vert}%
\def\bbra#1{\langle\!\langle {#1} \vert}%

\def\Tr{\operatorname{Tr}}
\def\transp#1{{#1}^\tau}

\def\Cmplx{\mathbb C}
\begin{document}
\begin{frontmatter}
\title{Optimal realization of the transposition maps}
\author{F. Buscemi, G. M. D'Ariano, P. Perinotti, and M. F. Sacchi}
\address{Theoretical Quantum
Optics Group\\ Universit\`a degli Studi di Pavia and INFM Unit\`a di
Pavia\\ via A. Bassi 6, I-27100 Pavia, Italy} 
\begin{abstract}
We solve the problem of achieving the optimal physical approximation
of the transposition for pure states of arbitrary quantum systems for
finite and infinite dimensions.  A unitary realization is also given
for any finite dimension, which provides the optimal quantum cloning
map of the ancilla as well.
\end{abstract}
\end{frontmatter}
\section{Introduction}
The formulation of no-go theorems in quantum information has given new
insight in the structure of the quantum theory itself.  The most
relevant examples are given by the no-cloning \cite{woot,yuen},
no-broadcasting \cite{nobr}, no-deleting \cite{pati} theorems, along
with the impossibility of measuring the wave-function of a single
system \cite{single}, and the debated quantum bit-commitment
\cite{nota}.

\par A transformation which is not allowed by quantum mechanics
naturally poses the problem of investigating about the best
approximation that one can ach\-ieve in principle. Remarkably, the
optimal approximation of a forbidden transformation may be related to
the optimal procedure to perform some information tasks. For example,
universal cloning is deeply related to the optimal eavesdropping in
cryptographic channels \cite{eav}. Moreover, one can infer from the
structure of the optimal approximation of unphysical maps a number of
properties about different topics in quantum mechanics, such as state
estimation \cite{chiara} and signaling \cite{Gisin}.

\par In this Letter we provide the optimal approximation of the
transposition for pure states both in finite and infinite dimensional
Hilbert spaces. Trans\-pos\-ition---in particular, {\em partial}
transposition---plays a major role in separability criteria for
bipartite states \cite{peres}. It is the simplest example of a
positive map, which is not completely positive. In fact, such a map is
antilinear and for qubits it is related to the more familiar $U$-not
transformation that maps any arbitrary state to its orthogonal. Of
course, for systems with higher dimension the $U$-not map is not
uniquely defined, because more than one state is orthogonal to a given
one, whereas the transposition map depends on a choice of basis.

\par The paper is organized as follows.  In Sec. II we derive the
optimal transposition map for pure states in terms of fidelity for the
case of arbitrary finite dimensional Hilbert spaces. The derivation is
obtained by means of group representation theory. In Sec. III we give
explicitly a unitary realization of the optimal map for any
dimension. Such a unitary realization provides also the optimal
quantum cloning from one to two copies.  This result proves that the
recent experimental scheme proposed in Ref. \cite{dema1}, where the
optimal $U$-not and two optimal clones are created for qubits
simultaneously by the same machine, can be generalized to quantum
systems with arbitrary dimension. Section III is devoted to the
continuous variables case. Similarly, as for the cloning map, one has
to restrict the covariance group under which the map is universal. We
consider the case of Weyl-Heisenberg group, which provides the optimal
transposition map for coherent and squeezed states.
\section{Optimal transposition for finite dimension}
It is well known that the transposition map
\begin{equation}
\rho\longmapsto\rho^T
\end{equation}
is not physical since it is not completely positive (CP). As stated in
the introduction the problem naturally arises to find the optimal
physical. i.e. CP map $\mathcal M$ whose output has
maximal fidelity  with the transposed input. 
We consider pure input states, for which the fidelity writes 
\begin{equation}
F=\Tr\left[{\rho^T}\mathcal M(\rho)\right]\,.
\label{eq:fid}
\end{equation}

\par We settle here some useful notation that will be used along the
paper. A generic vector in the bipartite Hilbert space $\mathcal
H\otimes\mathcal H$, with $\hbox{dim}(\mathcal H)=d$, can be expanded
on a fixed factorized orthonormal basis as
$\sum_{i,j=1}^d\Psi_{ij}\ket{i}\ket{j}$.  This naturally defines a one
to one correspondence between vectors in $\mathcal H\otimes\mathcal H$
and linear operators in $\mathcal L(\mathcal H)$. The operator $\Psi
=\sum_{ij}\Psi_{ij}\ket{i}\bra{j}$ can be used to label the state as
follows
\begin{equation}
\kket{\Psi}=\sum
_{i,j=1}^d\Psi_{ij}\ket{i}\ket{j}\,.
\end{equation}
In this framework one can easily verify the useful identities \cite{pla}
\begin{eqnarray}
&&A\otimes C \kket{B}=\kket{ABC^T}\;,
\nonumber\\
&&\Tr_1[\kket{A}\bbra{B}]=A^TB^* \;,\nonumber\\
&&\Tr_2[\kket{A}\bbra{B}]=AB^\dag\;,
\end{eqnarray}
where $T$ and $*$ denote transposition and complex conjugation with
respect to the orthonormal basis $\{| i \rangle \}$. 
The set of possible maps $\mathcal M$ can be parametrized using the 
isomorphism \cite{jami} between CP maps and positive operators
\begin{eqnarray}
&&R_\mathcal M =\mathcal M\otimes\mathcal I \kket{I}\bbra{I}\;,\nonumber\\
&&\mathcal M(\rho)=\Tr_2[(I\otimes\rho^T )R_\mathcal M]\;.
\label{eq:mapop}
\end{eqnarray}
Using Eq. (\ref{eq:mapop}), the expression for the fidelity writes 
\begin{equation}
F=\Tr[(\rho^T\otimes\rho^T )R_\mathcal M]\,,
\end{equation}
where $R_\mathcal M$ is a positive operator that satisfies the constraint
\begin{equation}
\Tr_1[R_\mathcal M]=I_2\;,
\label{eq:tracepr}
\end{equation}
because of the trace-preserving condition of the map $\mathcal M$.  
We pose the further constraint that $\mathcal M$ is
universal, namely that it is covariant under the action of the group
$\mathrm{SU}(d)$. 
On physical grounds this means that we restrict
attention to maps that equally well approximate the transposition,  
independently of the input pure state. The covariance property for the
transposition map
reads
\begin{equation}
\mathcal M(U\rho U^\dag)=U^*\mathcal M(\rho)U^T\,,\quad\forall \ 
U\in\mathrm{SU}(d)\,,
\end{equation}
and this is equivalent to the following condition on $R_\mathcal M$
\cite{clon}
\begin{equation}
(U^*\otimes U^*)R_\mathcal M(U^T\otimes U^T)=R_\mathcal M
\;.
\label{eq:covar}
\end{equation}
Since for $\mathrm{SU}(d)$ the representation $U^*\otimes U^*$ can be
decomposed into two inequivalent irreducible representations supported
by the totally symmetric and totally antisymmetric subspaces of
$\mathcal H\otimes\mathcal H$, according to Schur's lemma, condition
(\ref{eq:covar}) implies the following form for $R_\mathcal M$
\begin{equation}
R_\mathcal M=c_A P_A+c_s P_S\,.
\label{eq:covrm}
\end{equation}
The operators $P_S$ and $P_A$ are  the projectors on the
totally symmetric and totally antisymmetric spaces, respectively, and
can be written as 
\begin{eqnarray}
&&P_S=\frac12(I+E)\;,\nonumber \\& & 
P_A=\frac12(I-E)
\;,\label{}
\end{eqnarray}
where $E$ denotes the swap operator on $\mathcal H\otimes\mathcal H$, namely
$E\ket{\phi}\ket{\psi}=\ket{\psi}\ket{\phi}$ for all
$\phi,\psi\in\mathcal H$. The trace-preserving condition in
Eq. (\ref{eq:tracepr}), along with 
the positivity constraint rewrite
\begin{eqnarray}
&&c_S\frac{d+1}2+c_A\frac{d-1}2=1 \;,\nonumber\\
&&c_S,c_A\geq0\,.
\label{eq:constra}
\end{eqnarray}
Due to the covariance condition, the fidelity F  
is independent of the input state $\rho$. 
Using Eq. (\ref{eq:covrm}) one has 
\begin{equation}
F=c_S\Tr[(\rho^T\otimes\rho^T)
    P_S]+c_A\Tr[(\rho^T\otimes\rho^T )P_A]=c_S\,.
\end{equation}
Upon  maximizing $c_S$ with the 
constraints in Eq. (\ref{eq:constra}), the optimal map is obtained 
for 
\begin{equation}
c_S=\frac2{d+1}\,,\quad c_A=0\,.
\end{equation}
Correspondingly, one has $R_\mathcal M=\frac2{d+1}P_S$,  and hence 
\begin{equation}
\mathcal M(\rho)=\frac2{d+1}\Tr _2 [(I\otimes\rho^T )P_S]
=\frac1{d+1}(I+\rho^T)\;.\label{mapp}
\end{equation}
The optimal fidelity is then given by 
\begin{equation}
F=\frac2{d+1}\Tr[(\rho^T\otimes\rho^T )
  P_S]=\frac2{d+1} \,.
\end{equation}
The state in Eq. (\ref{mapp}) coincides with the anticlone state of
Ref. \cite{buzek} for the universal cloning machine from one to two
copies. Moreover, the map $\mathcal M$ is the same as the ``structural
physical approximation'' of the transposition of Ref. \cite{horod}.
Here, we proved the optimality of $\mathcal M$ without assumptions,
thus also showing that the anticlone corresponds to the optimal
transposed state.

\par A Kraus decomposition of the map $\mathcal M$ can be obtained by
diagonalizing $R_\mathcal M$ as follows
\begin{eqnarray}
&&R_\mathcal
  M=\frac1{d+1}(I+E)=\frac1{d+1}\sum_{mn}(\ket{m}\bra{m}\otimes\ket{n}\bra{n}+\ket{m}\bra{n}\otimes\ket{n}\bra{m})\nonumber\\
&&=\frac1{2(d+1)}\sum_{mn}(\kket{mn}+\kket{nm})(\bbra{mn}+\bbra{nm})=\sum_{mn}\kket{M^S_{mn}}\bbra{M^S_{mn}}\,,
\end{eqnarray}
where
$M^S_{mn}=\frac1{\sqrt{2(d+1)}}(\ket{m}\bra{n}+\ket{n}\bra{m})$. 
The Kraus decomposition is then given by  
\begin{equation}
{\mathcal M}(\rho )=\sum_{mn}M^S_{mn}\rho M^S_{mn}\,.
\end{equation}
\par A Stinespring form of the map $\mathcal M$ can be written for an
ancilla in the Hilbert space $\mathcal H^{\otimes2}$ as follows
\begin{equation}
\mathcal M(\rho)=\Tr_{23}[V\rho V^\dag]\,,
\end{equation}
where $V$ denotes the isometry
\begin{equation}
V=\sum_{mn}M^S_{mn}\otimes\kket{mn}_{23}\,.
\label{eq:iso}
\end{equation}
We can verify  that $V$  is also an isometric extension for the
realization of the optimal  
universal cloning from $1$ to $2$ copies. In fact, upon
tracing out the system $1$, one obtains the optimal cloning map in the
Werner expression \cite{ww}
\begin{equation}
\mathcal C (\rho )=
\Tr_{1}[V\rho V^\dag]=\frac2{d+1}P_{S_{23}}(I_2\otimes\rho_3 )P_{S_{23}}\,.
\label{pro}
\end{equation}
This result also shows that the recent experimental scheme
proposed in Ref. \cite{dema1}, where the optimal $U$-not and two
optimal clones are created for qubits simultaneously by the same
machine, can be generalized to quantum systems with arbitrary
dimension. Notice that the cloning map is basis independent, whereas
the transposition map depends on the choice of the basis, which is
reflected by the particular Stinespring extension.  
 
\par In the following we explicitly derive a  unitary realization 
of the optimal map in Eq. (\ref{mapp}). 

\section{Unitary realization}
The isometry in Eq. (\ref{eq:iso}) provides the optimal universal
transposition map or the optimal universal cloning from one to two
copies, by tracing over the ancilla spaces 2 and 3 or the input space
1, respectively.  Starting from the isometry $V$, we look for a
unitary interaction $U$ between the system and a fixed preparation of
the ancilla that dilates $V$. For the explicit construction of the
unitary dilation $U$, we will follow the general framework of
Ref. \cite{preparation}.

\par First, notice that we can rewrite Eq. (\ref{eq:iso}) as
\begin{eqnarray}
V&=&\sqrt{\frac{2}{d+1}}\sum_{m,n=0}^{d-1}\ket{m}\frac{\ket{m}\ket{n}+\ket{n}\ket{m}}{2}\bra{n}\nonumber\\
&=&\sqrt{\frac{2}{d+1}}\sum_{m,q=0}^{d-1}\ket{m}\frac{\ket{m}\ket{m\oplus
q}+\ket{m\oplus q}\ket{m}}{2}\bra{m\oplus q}\nonumber\\
&=&\sqrt{\frac{2}{d+1}}
\left(V_{0,0}+\frac{1}{\sqrt{2}}\sum_{q=1}^{d-1}V_{0,q}^{(S)}\right)\;,
\end{eqnarray}
where we defined the operators
\begin{eqnarray}
V_{p,p}&=&\sum_{k=0}^{d-1}\ket{k}\ket{k\oplus p}\ket{k\oplus
  p}\bra{k\oplus p}\;, \nonumber\\
  V_{p,q}^{(S)}&=&\frac{1}{\sqrt{2}}\sum_{k=0}^{d-1}\ket{k}(\ket{k\oplus
  p}\ket{k\oplus q}+\ket{k\oplus q}\ket{k\oplus p})\bra{k\oplus q}\;,
\end{eqnarray}
with $p\neq q$ and $p,q=0,\dots,d-1$. In order to construct a unitary
realization $U$, we also define 
\begin{equation}
V_{p,q}^{(A)}=\frac{1}{\sqrt{2}}\sum_{k=0}^{d-1}\ket{k}(\ket{k\oplus p}\ket{k\oplus q}-\ket{k\oplus q}\ket{k\oplus p})\bra{k\oplus q}\,.
\end{equation}
One can easily verify that 
\begin{eqnarray}
&&V_{p,p}^\dag V_{q,r}^{(S)}=V_{p,p}^\dag
  V_{s,t}^{(A)}=V_{q,r}^{(S)\dag}V_{s,t}^{(A)}=0\,,\qquad\forall\  p,q,r,s,t=0,\dots,d-1\nonumber\\
&&V_{p_1,p_1}^\dag V_{p_2,p_2}=\delta_{p_1,p_2}\
  I_\mathcal{H}\,,\qquad\forall  p_1,p_2 \;,
\nonumber\\
&&V_{q_1,r_1}^{(S)\dag}V_{q_2,r_2}^{(S)}=\delta_{q_1,q_2}\delta_{r_1,r_2}\
  I_\mathcal{H}\,,\qquad\forall\  q_1 < r_1, q_2 < r_2\;,\nonumber\\
&&V_{s_1,t_1}^{(A)\dag}V_{s_2,t_2}^{(A)}=\delta_{s1,t_2}\delta_{s_1,t_2}\
  I_\mathcal{H}\,,\qquad\forall\  s_1 < t_1, s_2 < t_2\;,
\end{eqnarray}
namely, the three sets $\{V_{p,p}\}$, $\{V_{p,q}^{(S)}\}_{p<q}$ and
$\{V_{p,q}^{(A)}\}_{p<q}$ are orthogonal sets of orthogonal
isometries. Hence, the following operator 
\begin{equation}\label{unitary}
U=\sum_{p=0}^{d-1}V_{p,p}\otimes\bra{p}\bra{p}+\sum_{{p,q=0\atop p<q}}^{d-1}V_{p,q}^{(S)}\otimes \frac{\bra{p}\bra{q}+\bra{q}\bra{p}}{\sqrt{2}}+\sum_{{p,q=0\atop p<q}}^{d-1}V_{p,q}^{(A)}\otimes \frac{\bra{p}\bra{q}-\bra{q}\bra{p}}{\sqrt{2}}\,
\end{equation}
satisfies the unitary condition
\begin{equation}
U^\dag U=UU^\dag=I_\mathcal{H}\otimes(P_{S_{23}}+P_{A_{23}})=I_{\mathcal{H}^{\otimes 3}}\,.
\end{equation}
The optimal universal transposition map can be obtained as follows 
\begin{equation}
\mathcal{M}(\rho)=\Tr_{2,3}[U (\rho\otimes\kket{\phi}\bbra{\phi}) U^\dag]\;,
\end{equation}
where $\kket{\phi}\in\mathcal{H}^{\otimes 2}$ is the fixed normalized
totally symmetric ancilla state
\begin{equation}
\kket{\phi}=\sqrt{\frac{2}{d+1}}P_{S_{23}}\sum_{r=0}^{d-1}\ket{0}_2\ket{r}_3\,.
\end{equation}
As noted above, the unitary  $U$ provides the optimal universal
cloning as well, namely one has 
\begin{equation}
\mathcal C (\rho)=\Tr_{1}[U( \rho\otimes\kket{\phi}\bbra{\phi}) U^\dag]\,.
\end{equation}
\par For qubits, i.e. $d=2$, we obtain the network model of Ref. 
\cite{network}, with
\begin{equation}
U=
\begin{pmatrix}
1 & 0 & 0 & 0 & 0 & 0 & 0 & 0\\
0 & 0 & 0 & 0 & 0 & 1 & 0 & 0\\
0 & 0 & 0 & 0 & 0 & 0 & 1 & 0\\
0 & 0 & 0 & 0 & 0 & 0 & 0 & 1\\
0 & 0 & 0 & 1 & 0 & 0 & 0 & 0\\
0 & 0 & 1 & 0 & 0 & 0 & 0 & 0\\
0 & 1 & 0 & 0 & 0 & 0 & 0 & 0\\
0 & 0 & 0 & 0 & 1 & 0 & 0 & 0\\
\end{pmatrix}\,,
\end{equation}
and
$\kket{\phi}=\frac{1}{\sqrt{6}}(2\ket{0}_2\ket{0}_3+\ket{0}_2\ket{1}_3+\ket{1}_2\ket{0}_3)$.

\section{Continuous variables optimal transposition}
In the limit of dimension $d\to\infty$, the fidelity $F$ for the
universal transposition map goes to zero. However, as regards infinite
dimensional systems one can look for transposition maps that are not
universal, but covariant just for a group with reduced symmetry. The
typical covariance group for infinite dimensional quantum systems is
the Weyl-Heisenberg group, in the representation of displacement
operators $D(\alpha)=\exp(\alpha a^\dag-\alpha^* a)$, with
$\alpha\in\mathbb C$, and $a$ and $a^\dag$ being the annihilation and
creation operators. The covariant transposition map acts with the same
fidelity over any state obtained from a given pure state by
application of the displacement operator with arbitrary amplitude.
Such a covariance condition reads
\begin{equation}\mathcal M(D(\alpha)\rho D^\dag(\alpha))=D^*(\alpha)
\mathcal M(\rho)D^T(\alpha)\,,\quad\forall\alpha\in\Cmplx\,,
\end{equation}
which rewrites for the operator $R_\mathcal M$ as follows
\begin{equation}
[D^*(\alpha)\otimes D^*(\alpha),R_\mathcal
  M]=0\,,\quad\forall\alpha\in\Cmplx\,.
\label{eq:comm}
\end{equation}
The operator $R_\mathcal M$ can be expanded on the basis of
displacement operators, which are a spanning set for 
Hilbert-Schmidt operators on $\mathcal H$, namely 
\begin{equation}
R_\mathcal M=\int_\mathbb C\frac{\mathrm d^2\alpha}{\pi}\int_\mathbb C\frac{\mathrm d^2\beta}{\pi}\,r(\alpha,\beta)D(\alpha)\otimes D(\beta)\,.
\end{equation}
The condition in Eq. (\ref{eq:comm}) is then equivalent to
\begin{eqnarray}
&&\int_\mathbb C\frac{\mathrm d^2\alpha}{\pi} \int_\mathbb
C\frac{\mathrm d^2\beta}{\pi}\,r(\alpha,\beta)\mathrm
e^{\gamma(\alpha+\beta)-\gamma^*(\alpha^*+\beta^*)}D(\alpha)\otimes
D(\beta)=\nonumber\\ &&\int_\mathbb C\frac{\mathrm
d^2\alpha}{\pi}\int_\mathbb C\frac{\mathrm
d^2\beta}{\pi}\,r(\alpha,\beta)D(\alpha)\otimes D(\beta)\,,
\quad\forall\gamma\in\Cmplx\,,
\end{eqnarray}
and this is possible for $r(\alpha,\beta)=\pi r(\alpha )\,
\delta^2(\alpha+\beta)$, with $r(\alpha )$ complex function of $\alpha
$, thus giving
\begin{equation}
R_\mathcal M=\int_\mathbb C\frac{\mathrm
d^2\alpha}{\pi}\,r(\alpha)D(\alpha)\otimes D^\dag(\alpha)\,.
\end{equation}
The trace-preserving condition  $\Tr_1[R_\mathcal M]=I_2$ corresponds
to $r(0)=1$. Upon introducing the $50/50$ beam splitter operator 
$V=\exp[\frac \pi 4 (a^\dag
b-ab^\dag)]$, we can write 
\begin{equation}
R_\mathcal M= V\int_\mathbb C\frac{\mathrm d^2\alpha}{\pi}
\,r(\alpha)\,[D(\sqrt{2}\alpha)\otimes I]\,V^\dag
\equiv \frac12V^\dag (\xi\otimes\openone )V\,,
\end{equation}
and $R_\mathcal M$ is positive if and only if the following operator 
\begin{equation}
\xi \equiv 
\int_\Cmplx\frac{\d^2\alpha}{\pi}\,r\left(\frac\alpha{\sqrt{2}}\right)D(\alpha)
\end{equation}
is positive. Since $r(0)=1$,  then $\hbox{Tr}[\xi ]=1$, namely
$\xi $  is a state. 
For the covariance condition in Eq. (\ref{eq:comm}), the fidelity of
the map with the state $\transp{[D(\alpha)\rho D^\dag(\alpha)]}$ is independent
of $\alpha$ and just depends on the seed
$\rho$. One has 
\begin{equation} 
F=\frac 12 \Tr[(\transp{\rho}\otimes\transp{\rho})V(\xi\otimes\openone)V^\dag]\,.
\label{fed}
\end{equation} 
Equation (\ref{fed}) is linear in $\xi$, which lies in a convex
set. The maximum fidelity is then achieved by a pure state
$\xi=|\chi \rangle \langle \chi|$, and the optimal  map is given by
\begin{equation}
R_\mathcal M=\frac12V(|\chi \rangle \langle \chi|\otimes\openone)V^\dag\,.
\end{equation}
The vector $|\chi \rangle $ can be determined as the eigenvector
corresponding to the maximum eigenvalue of the state
\begin{equation}
\Tr_2[V^\dag(\transp{\rho}\otimes\transp{\rho})V]\,.
\end{equation}
The explicit form of the map acting on a general state $\sigma $ is given by
\begin{equation}
\mathcal M (\sigma )=
\frac12\Tr_2[(\openone\otimes\transp{\sigma })V(|\chi \rangle \langle 
\chi|\otimes\openone) V^\dag]\;,
\end{equation}
and provides the optimal transposition for any pure state. We remind
that $|\chi \rangle $ depends on the seed state $\rho $. 
Notice that for coherent states, namely for $\rho=|0\rangle\langle0|$,
the optimal transposed state can be obtained as the anticlone from the
optimal covariant cloning from one to two copies \cite{cerf}, with
optimal fidelity $F=1/2$, 
generalizing the result for the finite dimensional case.

\section*{Acknowledgments}
This work has been sponsored by INFM through the project
PRA-2002-CLON, and by EEC and MIUR through the cosponsored ATESIT 
project IST-2000-29681 and Cofinanziamento 2002.

\end{document}